# Cluster Evaluation of Density Based Subspace Clustering

Rahmat Widia Sembiring, Jasni Mohamad Zain

**Abstract** – Clustering real world data often faced with curse of dimensionality, where real world data often consist of many dimensions. Multidimensional data clustering evaluation can be done through a density-based approach. Density approaches based on the paradigm introduced by DBSCAN clustering. In this approach, density of each object neighbours with MinPoints will be calculated. Cluster change will occur in accordance with changes in density of each object neighbours. The neighbours of each object typically determined using a distance function, for example the Euclidean distance. In this paper SUBCLU, FIRES and INSCY methods will be applied to clustering 6x1595 dimension synthetic datasets. IO Entropy, F1 Measure, coverage, accurate and time consumption used as evaluation performance parameters. Evaluation results showed SUBCLU method requires considerable time to process subspace clustering; however, its value coverage is better. Meanwhile INSCY method is better for accuracy comparing with two other methods, although consequence time calculation was longer.

**Index Terms** — clustering, density, subspace clustering, SUBCLU, FIRES, INSCY.

——————————— ◆ ———————————

## 1 DATA MINING AND CLUSTERING

Data mining is the process of extracting the data from large databases, used as technology to generate the required information. Data mining methods can be used to predict future data trends, estimate its scope, and can be used as a reliable basis in the decision making process. Functions of data mining are association, correlation, prediction, clustering, classification, analysis, trends, outliers and deviation analysis, and similarity and dissimilarity analysis.

One of frequently used data mining method to find patterns or groupings of data is clustering. Clustering is the division of data into objects that have similarities. Showing the data into smaller clusters to make the data becomes much simpler, however, can also be loss of important piece of data, therefore the cluster needs to be analyzed and evaluated.

This paper organized into a few sections. Section 2 will present cluster analysis. Section 3 presents density-based clustering, followed by density -based subspace clustering in Section 4. Our proposed experiment based on performance evaluation discussed in Section 5, followed by concluding remarks in Section 6.

———————————————

- *Rahmat Widia Sembiring, is with Faculty of Computer Systems and Software Engineering Universiti Malaysia Pahang, Lebuhraya Tun Razak, 26300, Kuantan, Pahang Darul Makmur, Malaysia, E-mail : rahmatws@yahoo.com*
- *Jasni Mohamad Zain, is Associate Professor at Faculty of Computer Systems and Software Engineering Universiti Malaysia Pahang, Lebuhraya Tun Razak, 26300, Kuantan, Pahang Darul Makmur, Malaysia, E-mail : jasni@ump.edu.my*

## 2 CLUSTER ANALYSIS

Cluster analysis is a quite popular method of discretizing the data [1]. Cluster analysis performed with multivariate statistics, identifies objects that have similarities and separate from the other object, so the variation between objects in a group smaller than the variation with objects in other groups.

Cluster analysis consists of several stages, beginning with the separation of objects into a cluster or group, followed by appropriate to interpret each characteristic value contained within their objects, and labelled of each group. The next stage is to validate the results of the cluster, using discriminant function.

## 3 DENSITY BASED CLUSTERING

Density-based clustering method calculating the distance to the nearest neighbour object, object measured with the objects of the local neighbourhood, if inter-object close relative with its neighbour said as normal object, and vice versa.

In density-based cluster, there are two points to be concerned; first is density-reachable, where $p$ point is density-reachable from point $q$ with *Eps*, *MinPoints* if there are rows of points $p_1, ..., p_n$, $p_1 = q$, $p_n = p$, such that $p_{i+1}$ directly density-reachable from $p_i$ as shown in Figure-1.

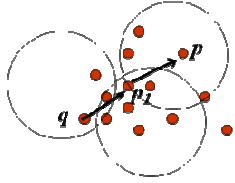

*Figure-1 Density Reachable*

The second point is density-connected, where point *p* as density-connected at the point *q* with *Eps*, *MinPoints* if there is a point *o* such that *p* and *q* density-reachable from *o* with *Eps* and *MinPoints*, as shown in Figure-2.

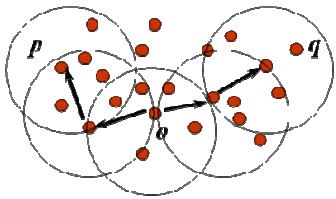

*Figure-2 Density Connected*

Density-based clustering algorithm will classify objects based on object-specific functions. The most popular algorithm is DBSCAN [2]. In DBSCAN, the data do not have enough distance to form a cluster, referred as an outlier, will be eliminated. DBSCAN will determine for themselves the number of clusters generated after inclusion of input *Eps* (maximum radius of neighbourhood of a point) and *MinPoints* (minimum number of points which is in an environment *Eps*), expressed in *pseudocode* algorithms such as Figure-3 [Wikipedia].

```
DBSCAN(D, eps, MinPts)
  C = 0
  for each unvisited point P in dataset D
    mark P as visited
    N = getNeighbours (P, eps)
    if sizeof(N) < MinPts
      mark P as NOISE
    else
      C = next cluster
      expandCluster(P, N, C, eps, MinPts)

expandCluster(P, N, C, eps, MinPts)
  add P to cluster C
  for each point P' in N
    if P' is not visited
      mark P' as visited
      N' = getNeighbours(P', eps)
      if sizeof(N') >= MinPts
        N = N joined with N'
      if P' is not yet member of any cluster
        add P' to cluster C
```

*Figure-3 DBSCAN Algorithm*

## 4 DENSITY BASED SUBSPACE CLUSTERING

Subspace clustering is a method to determine the clusters that form on a different subspaces, this method is better in handling multidimensional data than other methods.

Figure-4 (wikipedia) shows the two dimensions of the clusters placed in a different subspace. On the dimension of the subspace cluster $c_{a1}$ (in the subspace *{x}*) and $c_b$, $c_c$, $c_d$ (in the subspace *{y}*) can found. Meanwhile $c_c$ not included in the subspace cluster. In two-dimensional cluster, $c_{ab}$ and $c_{ad}$ identified as clusters.

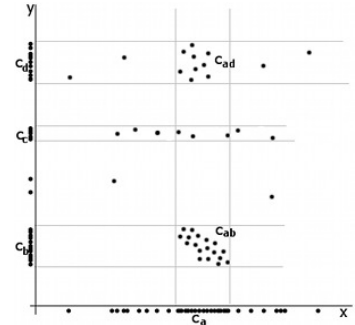

*Figure-4 Subspace clustering*

There is some discussion of subspace clustering ([3], [4], [5], [6], [6], [8], [9], [10], [11], [12], [13], [14], [15], [16], [17], [18], [19]). The main problem in clustering is the cluster can be in a different subspace, with the combination of different dimensions, if the number of dimensions higher, caused more difficult to find clusters. Subspace clustering method will automatically find the units clustered in each subspace. As clustering in general, important to analyze in subspace clustering is the problem of density of each data object. In this paper will discuss the application of SUBCLU [20], FIRES [21], and INSCY [22] for subspace clustering.

SUBCLU (density-connected *SUBspace CLUstering*) [20], is an effective and efficient method in subspace clustering problems. Using the concept of density in relation DBSCAN [2], with grid-based approach, this method can detect the shape and position of clusters in the subspace. Monotonous nature of the relationship density, bottom-up approach used to pruning subspace and produces clusters that are connected with density (Figure-5) and which are not connected with the density (Figure-6).

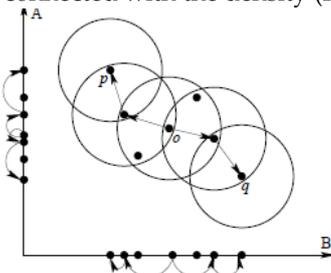

*Figure-5 Cluster with density relation*

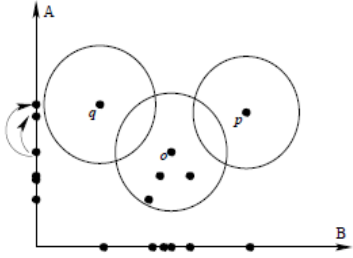

*Figure-6 Cluster without density relation*

As long as no unnecessary subspace produced, the result will be the same SUBCLU obtained by DBSCAN, SUBCLU processes can be seen through the algorithm in Figure-7.

```
SUBCLU(SetOfObjects DB, Real ", Integer m)
    /* STEP 1 Generate all 1-D clusters */
    S1 := ; // set of 1-D subspaces containing clusters
    C1 := ; // set of all sets of clusters in 1-D subspaces
    FOR each ai 2 A DO
        Cfaig := DBSCAN(DB; faig; ";m) // set of all clusters in subspace ai;
        IF Cfaig 6= ; THEN // at least one cluster in subspace faig found
            S1 := S1 [ faig;
            C1 := C1 [ Cfaig;
        END IF
    END FOR
    /* STEP 2 Generate (k + 1)-D clusters from k-D clusters */
    k := 1;
    WHILE Ck 6= ;
        /* STEP 2.1 Generate (k + 1)-dimensional candidate subspaces */
        CandSk+1 := GenerateCandidateSubspaces(Sk);
        /* STEP 2.2 Test candidates and generate (k + 1)-dimensional clusters */
        FOR EACH cand 2 CandSk+1 DO
            // Search k-dim subspace of cand with minimal number of objects in the clusters
            bestSubspace := min
            s2Sk^s_cand
            P
            Ci2Cs
            jCij
            Ccand := ;;
            FOR EACH cluster cl 2 CbestSubspace DO
                Ccand = Ccand [ DBSCAN(cl; cand; ";m);
                IF Ccand 6= ; THEN
                    Sk+1 := Sk+1 [ cand;
                    Ck+1 := Ck+1 [ Ccand;
                END IF
            END FOR
        END FOR
        k := k + 1
    END WHILE
```

*Figure-7 SUBCLU Algorithm*

Second subspace method used in this paper is FIRES [21]. FIRES (*FIlter REfinement Subspace clustering*) framework based on efficiency filter refinement, by determining the frequency scale quadratic of data dimensions and dimensional subspace clusters. This method can applied to clusters that recognized based on the local threshold density.

FIRES consist of three phases, namely pre-clustering, in this phase all the so-called cluster *1D-base* clusters will be calculated, which can use existing methods of subspace clustering. The second phase is the generation of subspace cluster approximations, in this phase the existing clusters will combined to find the maximum dimensional subspace cluster approach, but not incorporate in apriori style, but using the scale most quadratic of the number of dimensions. The final stage is to post-processing subspace clusters, by smoothing the cluster on the second phase.

Another method that was used INSCY [22] (*INdexing Subspace Clusters with in-process-removal of redundancY*), used breadth first approach, by performing recursive mining in all parts of the cluster of subspace projections, before passing to the next section.

This strategy has two advantages, high dimensional maximal projection will done first, then perform pruning of all loop dimensions and gain efficiency, second, indexing potential of subspace clusters that may occur. For more details can see in the algorithm INSCY in Figure-8.

*foreach descriptor in scy-tree do*
*restricted-tree := restrict(scy-tree, descriptor);*
*restricted-tree := mergeWithNeighbors(restricted-tree);*
*pruneRecursion(restricted-tree); //prune sparse regions*
*INSCY(restricted-tree,result); //depth-first via recursion*
*pruneRedundancy(restricted-tree); //in-process-removal*
*result := DBClustering(restricted-tree) ∪ œsult;*

*Figure-8 INSCY Algorithm*

## 5 PERFORMANCE EVALUATION

We have tested SUBCLU, FIRES and INSCY using data sets, with processor 1.66 GHz and 1 GB of RAM.

### 5.1 Data Sets

In our experiments, synthetic data was used consisting of 6x1595 dimensions (6 attributes and 1595 instant / data) that our adoption of [12], graphically the initial data are as in Figure-9.

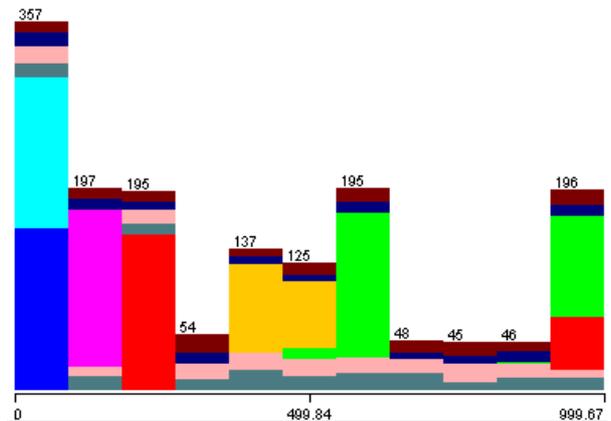

*Figure-9 Dataset Distribution*

Based on experiment with SUBCLU, FIRES and IN-SCY methods with some parameters, we obtained the results as in Table-1.

*Table-1 Experiment Result Performance*

| Method | No of cluster | Clustering time (ms) | Accuracy | Coverage | IO Entropy | F1 Measure | Calculation Time (ms) |
|---|---|---|---|---|---|---|---|
| SUBCLU | 31 | 58449 | 0,1 | 1 | 0,01 | 0,01 | 133 |
| FIRES | 2 | 569 | 0,1 | 0,01 | 0,64 | 0,01 | 14 |
| INSCY | 1 | 1899 | 0,39 | 0,34 | 1 | 0,42 | 1281 |

## 5.2 Efficiency

From the experimental results, we can see that the clustering time for SUBCLU method is more than 20 times longer than the FIRES and INSCY, as shown in Figure-10.

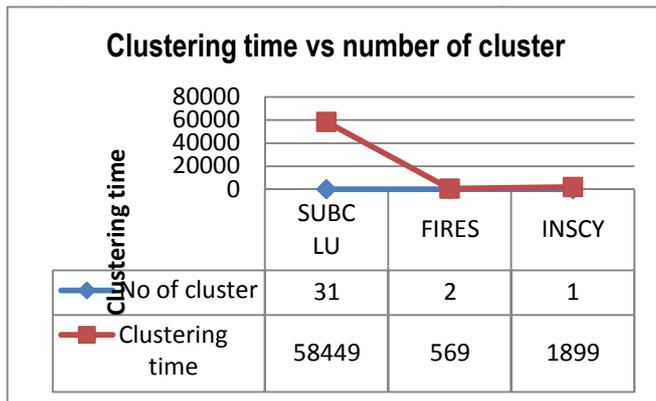

*Figure-10 Clustering time vs number of cluster*

Meanwhile, if evaluated based on the time required to perform calculations of each parameter, it can be seen that INSCY method requires a longer time than SUBCLU and FIRES methods, as shown in Figure-11.

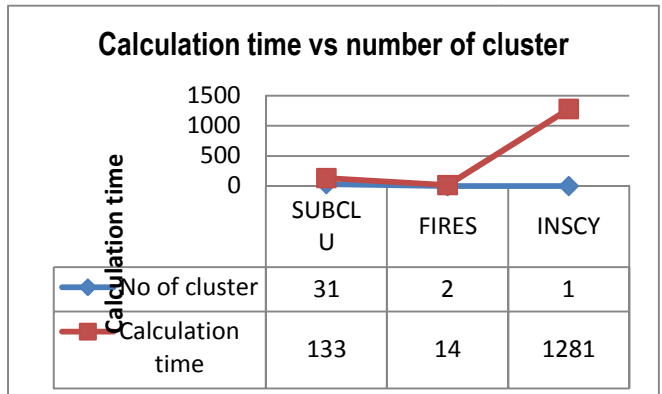

*Figure-11 Calculation time vs number of cluster*

## 5.3 Accurate

In addition to evaluating the work efficiency of sub-space clustering method, this paper also discusses other related parameters of clustering results. There are four parameters: accuracy, coverage, IO Entropy and F1 Entropy.

The experimental results show that the accuracy of INSCY method is more accurate than SUBCLU and FIRES, as shown in Figure-12.

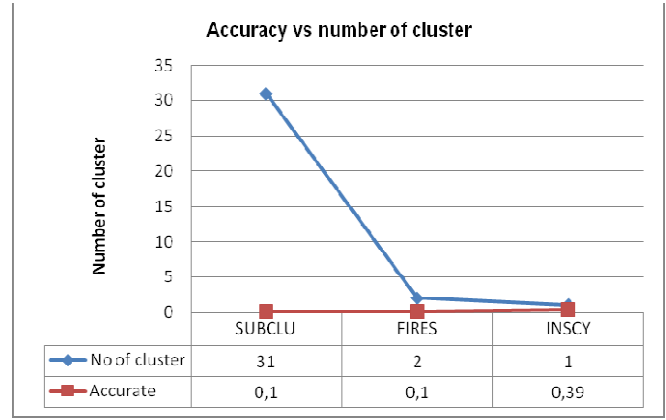

*Figure-12 Accuracy vs number of cluster*

IO Entropy used to evaluate the purity of the clustering, while the coverage used to evaluate the scope of the size of clustering. For coverage, SUBCLU method is better than FIRES and INSCY, as shown in Figure-13.

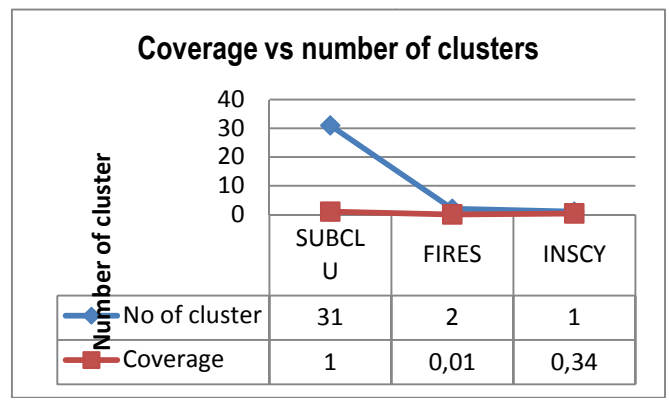

*Figure-13 Coverage vs number of cluster*

For IO Entropy, INSCY method is better than FIRES and SUBCLU method, as shown in Figure-14.

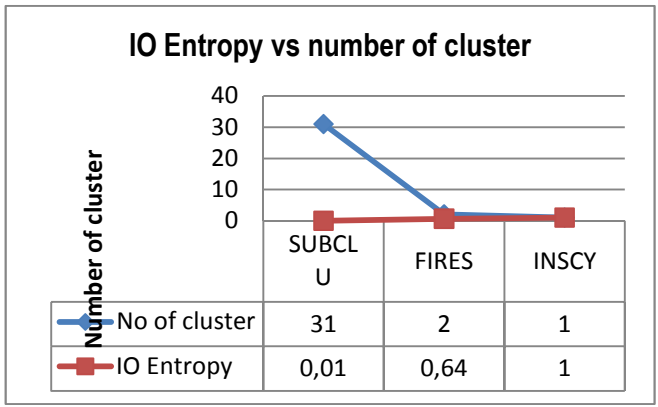

*Figure-14 IO Entropy vs number of cluster*

F1-Measure generally used to evaluate classifier, but also can be used to evaluate or projected subspace clustering, by measuring the average value of harmony from the cluster, whether all the clusters detected and precision (if all the clusters detected with accuracy). For F1 Measure INSCY method is better than FIRES and SUBCLU method, as shown in Figure-15.

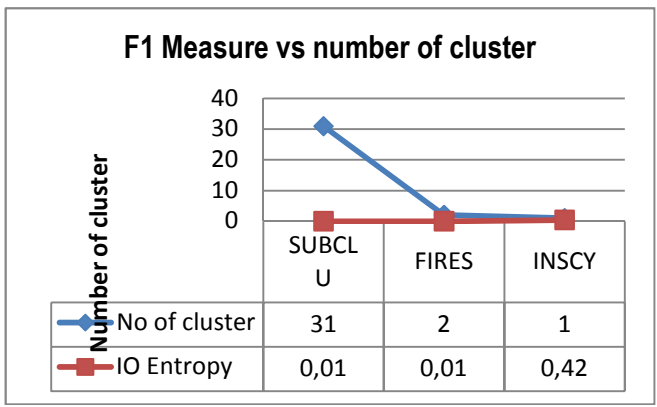

*Figure-15 F1 Measure vs number of clusters*

## 6 CONCLUSION

SUBCLU, FIRES and INSCY methods were applied to clustering 6x1595-dimension synthetic datasets. IO Entropy, F1 Measure, coverage, accurate and time consumption used as evaluation performance parameters. Evaluation results showed SUBCLU method assessed require however considerable time to process subspace clustering, its value coverage is better. Meanwhile INSCY method is better for accuracy comparing with two other methods, although consequence time calculation was longer. Research in subspace clustering method has a lot of potential to be developed further in the future. We will be conducting more in-depth study related to pre-processing, dimension reduction, and outlier detection of subspace clustering method in the future.

## ACKNOWLEDGMENT

The authors wish to thank Universiti Malaysia Pahang. This work supported in part by a grant from GRS090116.

**Rahmat Widia Sembiring** received the degree in 1989 from Universitas Sumatera Utara, Indonesia, Master degree in computer science/information technology in 2003 from Universiti Sains Malaysia, Malaysia. He is currently as Ph.D student in the Faculty of Computer Systems and Software Engineering, Universiti Malaysia Pahang. His research interests include data mining, data clustering, and database.

**Jasni Mohamad Zain** received the BSc.(Hons) Computer Science in 1989 from the University of Liverpool, England, UK, Master degree in 1998 from Hull University, UK, awarded Ph.D in 2005 from Brunel University, West London, UK. She is now an Associate Professor of Universiti Malaysia Pahang, also as Dean of Faculty of Computer System and Software Engineering, Universiti Malaysia Pahang.